FRONT MATTER

# Unusual strain relaxation and Dirac semimetallic behavior in epitaxial antiperovskite nitrides

**Authors:** Ting Cui,[1,2,†] Zihan Xu,[3†] Qinghua Zhang,[1,†] Xiaodong Zhang,[3] Qianying Wang,[1,2] Dongke Rong,[1,2] Songhee Choi,[1] Axin Xie,[1,2] Hongyun Ji,[1,2] Can Wang,[1,2] Chen Ge,[1,2] Hongjian Feng,[3] Shanmin Wang,[5] Kuijuan Jin,[1,2,*] Liang Si,[3,4,*] and Er-Jia Guo[1,2,*]

**Affiliations:**

[1] Beijing National Laboratory for Condensed Matter Physics and Institute of Physics, Chinese Academy of Sciences, Beijing 100190, China.

[2] Department of Physics & Center of Materials Science and Optoelectronics Engineering, University of Chinese Academy of Sciences, Beijing 100049, China.

[3] School of Physics, Northwest University, Xi'an 710127, China.

[4] Shaanxi Key Laboratory for Theoretical Physics Frontiers, Xi'an 710127, China.

[5] Department of Physics, Southern University of Science and Technology, Shenzhen 518055, China.

[†] These authors contributed equally to this work.

[*] Corresponding author Emails: kjjin@iphy.ac.cn, siliang@nwu.edu.cn, and ejguo@iphy.ac.cn

## Abstract

Antiperovskite nitrides ($X_3AN$) are the structural analogues to perovskite oxides, while their epitaxial growth and electronic properties remain largely unexplored. We report the successful synthesis of $Ni_3InN$ thin films on substrates with different lattice constants. First-principles phonon calculations confirm the dynamical stability of cubic phase $Ni_3InN$, providing the basis for epitaxial synthesis. High-resolution scanning transmission electron microscopy reveals coherent (001)-oriented interfaces when $Ni_3InN$ is grown on $LaAlO_3$ and $SrTiO_3$, while an unexpected (011)-orientation forms on $DyScO_3$, aligning with surface-energy predictions. Transport measurements highlight a strain-controlled Fermi-liquid behavior, correlated with variations in the Ni-3*d* bandwidth and hybridization. Band structure calculations reveal a dual character near the Fermi level: a high-mobility Dirac-like band and a Ni-3*d* manifold that drives strange-metal transport with a reduced slope compared to oxide perovskites. The formal Ni valence (~+2/3) places $Ni_3InN$ in an overdoped correlated-metal regime, distinguishing from most perovskite oxides. This positions antiperovskite nitrides as a promising platform for investigating overdoped Fermi liquids and strange-metal behavior.

## Teaser

Antiperovskite nitrides exhibit unexpected strain accommodation and Dirac-like transport behavior, demonstrating a promising candidate for investigating complex electron interactions.

## Main text

The search for emergent quantum states in correlated electron systems continues to be at the forefront of condensed matter physics. Many of these states—such as unconventional superconductivity (*1,2*), colossal magnetoresistance (*3,4*), and Fermi-liquid transport (*5,6*)—emerge from the complex coupling between various physical degrees of freedom, including charge, spin, orbital, and lattice. Perovskite oxides, with the general formula $ABO_3$ (Fig. 1A), have long served as a canonical platform for exploring these phenomena. For example, interface engineering in $ABO_3/A'B'O_3$ heterostructures has facilitated the discovery of two-dimensional electron gases and their associated superconductivity (*7,8*). The synthesis of thin films along unconventional crystallographic orientations (e.g., [110] or [111]) has led to the revelation of novel orbital and topological states (*9,10*). Additionally, (original and reduced) Ruddlesden–Popper (RP) phases, such as $La_3Ni_2O_7$, and $La_4Ni_3O_{10}$, have recently uncovered unconventional high-temperature superconductivity (*11-14*). More strikingly, topotactic reduction from $ABO_3$ to $ABO_2$ phases has stabilized new electronic ground states, as exemplified by superconductivity in the hole-doped infinite-layer nickelates (Nd,Sr)$NiO_2$ (*15,16*). These advances underscore how structural flexibility, heterostructure construction, and symmetry breaking within the perovskite framework can provide a rich platform for emergent phases.

Inspired by these advances, a key direction is to explore material families that share structural motifs with perovskites but have been comparatively less studied. Unlike their perovskite counterparts, antiperovskites with general formular of $A_3BX$ (X = O or N) feature inverted positions of cations and anions (Fig. 1B). However, this inversion is difficult to achieve in oxides due to the small ionic size and high electronegativity of $O^{2-}$ anion. These limitations lead to the structural and electrostatic instabilities, as well as the challenges in maintaining charge balance with suitable cation species (*17*). Fig. 1C presents the development timeline of both antiperovskite oxides and nitrides. Notably, only a limited number of antiperovskite oxides, including $Sr_3SnO$, $Ca_3SnO$, and $Sr_3PbO$, have been successfully stabilized and reported to date (*18-20*). The magnetism, thermoelectric, and superconductivity had been investigated in these antiperovskite oxides over decades. Compared to oxides, nitrides with antiperovskite structure are more readily formed (*21-24*). It preserves the favorable electrostatic and geometric coordination with low-valence metal cations. The partial covalency of metal–nitrogen bonds and the availability of suitable synthetic routes make nitrides more structurally and chemically compatible with the antiperovskite framework. Despite their simple cubic symmetry, anti-perovskite nitrides host a wide variety of intriguing states (*25-28*), ranging from negative thermal expansion in $Mn_3ZnN$, electron-phonon mediated superconductivity in $Ni_3ZnN$ (*21*), to exotic topological band structures in $Ca_3BiN$ and related compounds (*29*). $Mn_3GaN$, firstly predicted in 1964, then had been studied for their non-collinear antiferromagnetism and large anomalous Hall effects, which are highly relevant for spintronic applications (*28,30-32*). From practical application point of view, the negative thermal expansion behavior in antiperovskite nitrides can be useful for thermal management and as components in composition to compensate for expansion. The excellent mechanical and thermodynamic stability make them benefit for device integration (*33*). Thus, these advantages make them a compelling complementary system to conventional perovskite oxides. Yet, compared with the vast body of work on perovskite oxides, the epitaxial growth, interface engineering, and correlated-electron properties of antiperovskite nitrides remain largely unexplored.

In this work, we took $Ni_3InN$ as a model system to demonstrate its epitaxial thin film growth and strain-mediated physical properties of a dynamically stable antiperovskite nitride. We revealed distinct strain relaxation behavior that contrasts sharply with conventional perovskite oxides. This behavior arises from a combination of its high structural rigidity and reduced surface formation energy of the (011) facet under tensile strain. We found that the strain triggers the transition in the low-temperature transport behavior, which correlates with changes in Ni-3*d* bandwidth and Ni-3*d*/N-2*p* hybridization. Importantly, the formal Ni valence in $Ni_3InN$ is approximately +2/3, placing the system in a heavily over-doped regime that is rarely accessible in perovskite oxides, where transition-metal valences are typically closer to integer values. This renders anti-perovskite $Ni_3InN$ a natural platform for exploring over-doped correlated metals and testing theoretical models of Fermi-liquid stability.

**Results**

**Synthesis and characterization of antiperovskite nitride $Ni_3InN$ thin films**

To confirm the stability and well-defined lattice structure of indium-based antiperovskite nitrides, $X_3InN$ ($X$ = 3*d* elements), we firstly calculated the phonon spectra of all indium-based antiperovskite nitrides (Fig. 1D) containing 3*d* elements (for computational details see Method). The results show that the spectra of $Ti_3InN$, $Cr_3InN$, $Mn_3InN$, $Fe_3InN$, and $Ni_3InN$ show a complete absence of imaginary modes throughout the Brillouin zone, confirming their dynamical stability, whereas $V_3InN$ and $Co_3InN$ exhibit dynamical instabilities with imaginary frequencies in their phonon spectra. For $Ni_3InN$, the phonon density of states further reveals that the lighter N atoms dominate the high-frequency modes (around 16 THz), whereas the heavier Ni and In atoms, coupled with N, govern the low-frequency region below 8 THz (fig. S1). Such a distribution is characteristic of intermetallic systems, in which N atoms are embedded within a covalently bonded In–Ni sublattice. This finding not only validates the dynamical stability of $Ni_3InN$ but also underscores its strong potential for experimental synthesis and functional applications.

To synthesis the high-quality $Ni_3InN$, we firstly fabricated a stoichiometric $Ni_3InN$ ceramic target using a high-pressure and high-temperature synthesis approach. The XRD results identify its correct crystallinity and pure phase (fig. S2). Subsequently, $Ni_3InN$ thin films were grown on $LaAlO_3$, $SrTiO_3$, and $DyScO_3$ substrates via pulsed laser deposition (PLD) (see Method). The calculated misfit strains $\varepsilon = (a_{sub} - a_{film})/a_{film} \times 100\%$, are $\varepsilon = -1.30\%$, 1.69%, 2.81% for $LaAlO_3$, $SrTiO_3$, and $DyScO_3$ substrates, respectively. Interestingly, X-ray diffraction $\theta$-$2\theta$ scans (Fig. 2A) reveals that $Ni_3InN$ films grown on $LaAlO_3$ and $SrTiO_3$ substrates exhibit a (001)-oriented phase, whereas the $Ni_3InN$ film on $DyScO_3$ substrate shows an unconventional (011) orientation. To further investigate the atomic-scale structure of representative films, we performed high-resolution scanning transmission electron microscopy (STEM) imaging on these two $Ni_3InN$ thin films. Fig. 2B presents a representative high-angle annular dark-field (HAADF) image of a (001)-oriented $Ni_3InN$ film grown on $SrTiO_3$ with the inset showing the corresponding unit cell. Fig. 2C displays a similar HAADF-STEM image of a (011)-oriented $Ni_3InN$ film on $DyScO_3$, with the inset providing a magnified view of the unit cell. Both images unambiguously confirm the excellent crystallinity of $Ni_3InN$ films grown on different substrates, as evidenced by the well-defined atomic arrangements, minimal disorder, and sharp diffraction contrast in the HAADF images. These findings stand in stark contrast to conventional perovskite oxide films, which are epitaxially grown on substrates that induce different strain states (fig. S3). For perovskite oxides, the out-of-plane lattice constants typically vary systematically with in-plane strain, resulting in elastic structural deformations. However, the growth orientation of $Ni_3InN$ films is strongly influenced by the strain state of the underlying substrates. Fig. 2D illustrates the (001)-oriented growth of $Ni_3InN$ under compressive strain ($\varepsilon < 0$), where the film remains constrained by the substrate. In contrast, when grown on substrates with significant tensile strain, the film undergoes strain relaxation, leading to a preferential [011] orientation (Fig. 2E).

In the thin film growth, the substrates' induced strain strongly influences which crystallographic orientation is more energetically favorable. To analyze this mechanism, the surface formation energy ($E_{surf}$) was calculated for two possible orientations, [001] and [011], of $Ni_3InN$ films. The surface formation energy essentially measures the energetic cost of creating and maintaining a given crystallographic surface under specific strain conditions. The orientation with the lower surface formation energy is thermodynamically more stable and thus more likely to occur experimentally. As shown in Fig. 2F, under compressive strain and moderate tensile strain, the [001] orientation has lower $E_{surf}$. The films naturally stabilize in this orientation when grown on substrates imposing these strain states (e.g. $LaAlO_3$ and $SrTiO_3$). Under significant tensile strain, however, the [011] orientation becomes energetically more favorable than the (001). This indicates that the crystal structure of $Ni_3InN$ relaxes more effectively in the [011] orientation when subjected to strong tensile strain (e.g. $DyScO_3$). Importantly, this theoretical explanation matches our experimental observations. Thereby, these calculations provide strong evidence that the unconventional orientation transition is driven by strain-mediated energetics, establishing a clear mechanism for controlling crystalline orientation in antiperovskite nitride films.

On the other hand, we also calculated the mechanical properties of bulk $Ni_3InN$ and extended the comparative analysis to other $X_3InN$ compounds. Fig 2G presents the calculated elastic constants $C_{11}$, $C_{12}$, and $C_{44}$ for bulk $X_3InN$, obtained via using density-functional theory (DFT). All $X_3InN$ compounds exhibit enhanced $C_{11}$, reflecting strong resistance to uniaxial compression, while their $C_{44}$ values are significantly reduced, indicating diminished shear rigidity. From a structural perspective, this behavior arises from the robust covalent interactions in the In-N and $X$-N bonding networks, which dominate the longitudinal response. In contrast, shear deformations preferentially activate lower-energy octahedral tilting or rotational modes, a hallmark of perovskite-type frameworks (*34,35*). To further quantify these characteristics, the Young's modulus of $Ni_3InN$ was directly compared with those of conventional perovskite oxides. (Fig. 2H and Table I) (*36-43*). The inset illustrates the strain state of an antiperovskite film grown on a single-crystalline substrate, where $\sigma^R$ denotes equibiaxial tensile stress. Notably, $Ni_3InN$ exhibits a remarkably high Young's modulus (~201.5 GPa), underscoring its capability to not only match but potentially surpass conventional perovskite oxides in mechanical properties. In additional, we find that the substantial $C_{11}$ value suggests high speed of sound and superior thermal conductivity due to strong bond stiffness, confirming the large stiffness of $Ni_3InN$. Furthermore, $Ni_3InN$ exhibits a Poisson's ratio of 0.33, closely comparable to that of stainless steel (0.30–0.31) and characteristic of ionic–covalent solids and metallic alloys (*44-49*). This correspondence indicates that $Ni_3InN$ can maintain mechanical stability under both in-plane and hydrostatic stress conditions.

**Microstructure and epitaxial relationship of $Ni_3InN$/perovskite oxides heterostructures**

The unique strain relaxation behavior suggests that the interface structure of $Ni_3InN$ may be established early in the thin film growth process. To further probe the atomic-scale interface structure and epitaxial relationships of $Ni_3InN$ films on various substrates, we employed a combination of high-resolution STEM and electron energy-loss spectroscopy (EELS). Low-magnification HAADF-STEM images for $Ni_3InN$ films were provided in fig. S4. In addition, the interfacial atomic configurations and corresponding intensity line profiles along the growth direction for three $Ni_3InN$ films were compared in fig. S5. The compressively strained $Ni_3InN$ films exhibit coherent epitaxy along the [001] direction with a characteristic $a/2$ in-plane lattice offset and an ultrathin transition layer with a thickness of approximate 0.24 nm (Fig 3A). The atomic resolution EELS analysis demonstrates an InNi-LaO interfacial configuration (Fig 3D) to compensate the interfacial charge imbalance. The [001] orientation persists under moderate tensile strain ($\varepsilon = 1.69\%$) despite an increased interfacial gap distance of approximate 0.76 nm (Fig. 3B). The interface termination changes to a distinct InNi-$TiO_2$ (Fig. 3E). We further observed that the unit cells of $Ni_3InN$ undergo periodic lateral realignment with a characteristic period of ~11.7 nm, corresponding to 30 unit cells of $SrTiO_3$ (fig. S6). This periodicity can be quantitatively accounted for by the lattice constant mismatch between $Ni_3InN$ and the substrate. Further increasing the tensile strain induces a complete orientation transition to [011], accompanied by interfacial gap distance expansion to 1.05 nm (Fig. 3C). Fig. 3F summarizes a strong correlation between interfacial gap distance and misfit strain, with gap expansion from 0.24 to 1.05 nm as strain increases from -1.3% (compressive) to +2.81% (tensile). The $c/a$ ratios derived from HADDF-STEM images consistently approach unity ($c/a \approx 1$), unambiguously demonstrating the exceptional structural stability of $Ni_3InN$ under various strain states. The strain distributions within $Ni_3InN$ films grown on three different substrates were analyzed along both the in-plane and out-of-plane directions (figs. S7–S9). The results reveal that strain relaxation occurs locally at the heterointerface on the nanometer scale.

The narrow strain accommodation window and abrupt structural transitions contrast sharply with the gradual interfacial evolution observed in perovskites oxide. In perovskite oxides, lattice mismatch with the substrate is typically accommodated through continuous elastic deformation of the lattice, enabled by the flexibility of corner-sharing octahedral frameworks that can tilt and rotate over a wide strain range. In some cases, the substrate induced epitaxial strain can persist over hundred nanometers (*50,51*). In contrast, the $Ni_3InN$ films do not rely on gradual lattice distortion to relieve strain. Instead, once a critical strain threshold is exceeded, the system undergoes abrupt relaxation processes, such as an orientation transition from (001) to (011), reflecting the more rigid bonding environment of the intermetallic In–Ni sublattice in which N atoms are embedded. We attribute this anomalous strain response in $Ni_3InN$ likely stems from its exceptional mechanical stiffness, revealing a new paradigm for strain engineering in hard, covalent antiperovskite systems.

**Exceptional electronic stability of strained $Ni_3InN$ films**

To explore the impact of strain on the transport and optical properties of $Ni_3InN$, we employed DFT to compute the electronic band structure, density of states (DOS), Fermi surface and edge state (fig. S10). As shown in Fig. 4A and fig. S10, the band structure reveals that empty In-*p* and fully occupied N-*p* orbitals correspond to an $In^{1+}/N^{3-}$ configuration, leading to a nominal Ni valence of +⅔. This unusual low valence of Ni, stabilized by the antiperovskite lattice, may underlie moderate correlations, emergent magnetic or electronic behaviors distinct from conventional $ANiO_X$ and nickelate materials. Without spin-orbit coupling (SOC), $Ni_3InN$ exhibits a closed bandgap, with the bands crossing the Fermi level along R–X and near the M point, indicating metallic behavior. Notably, band inversion and two Dirac-like nodes appear near the M point and Fermi level (Figs. 4A and 4B). Including SOC opens a hybrid bandgap in the inverted region via N-*p* and Ni-*d* orbital hybridization. Tight-binding calculations confirm the emergence of boundary states (fig. S10), suggesting $Ni_3InN$ supports a quantum-spin-Hall-like (QSH) phase that are typically linked to high conductivity and mobility, as seen in topological systems like graphene and semimetals (*52,53*). The calculated 3D band structure (Fig. 4B) and Fermi surface (fig. S10) shows an anisotropic character, with distinct electron pockets and critical points along different crystallographic directions, highlighting the interplay between symmetry and transport anisotropy.

Temperature-dependent resistivity ($\rho$) measurements on three strained $Ni_3InN$ thin films (Fig. 4C) reveal nearly constant $\rho$ across the full temperature range, with only a slight upturn below 20 K, indicating Ferm-liquid transport character. The high-temperature resistivity is exceptionally stable, with a slope of ~$2.02\times10^{-8}$ Ω·cm/K, outperforming most reported antiperovskite nitrides (*25,54-56*). Our transport calculations (fig. S11) align well with experimental trends and low-temperature behavior. $\theta$–$2\theta$ scans (fig. S12) confirm structural stability from 10 to 300 K, showing only minor thermal expansion along the c-axis. Hall measurements (fig. S13) indicate that carrier density ($n$) in films on $LaAlO_3$ remains stable at $(5.69–6.49)\times10^{22}$ cm$^{-3}$, with mobility ($\mu$) nearly temperature-independent (fig. S14). Under compressive strain, $n$ peaks while $\mu$ is minimized; tensile strain reduces $n$ and enhances $\mu$. On $SrTiO_3$ and $DyScO_3$ substrates, $n$ decreases and $\mu$ increases with temperature due to stronger correlation in Ni-3*d*. These findings suggest that the combination of structural robustness and Dirac metallicity drives the remarkable resistivity stability—an intrinsic property of $Ni_3InN$ thin films. High-temperature $\rho$–$T$ curves show linear behavior, indicating conduction dominated by Ni-3*d* bands and "*strange metal*" characteristics. Under increasing tensile strain, the $\rho$–$T$ slopes increase slightly but remain far below those in oxide perovskites (Fig. 4C). The low Ni valence (~+⅔) places $Ni_3InN$ in an overdoped correlated-metal regime. At low temperatures (20–65 K), $\rho$(T) fits well to $\rho(T) = \rho_0 + AT^2$, with $\rho_0 \approx 1.55\times10^{-4}$ Ω·cm and $A \approx 2.27\times10^{-10}$ Ω·cm/K$^2$, consistent with Fermi liquid behavior. Under tensile strain, reduced Ni-*d* bandwidth enhances correlations, yielding more pronounced Fermi liquid signatures, particularly on DSO substrates. Optical conductivity measurements confirm multiple interband excitations. In addition to a dominant Drude peak ($t_{2g}$–$t_{2g}$ transitions), four peaks arise from $e_g$–$t_{2g}$, $t_{2g}$–In-5*p*, $e_g$–In-5*p*, and N-2*p*–$t_{2g}$ transitions, consistent with band structure and DOS calculations.

Electronic structure analysis reveals two main Fermi-level contributions: temperature-independent

QSH-like edge states and bulk states near the R–X point that are temperature-dependent. At high temperatures, all films exhibit "*strange metal*" behavior with smaller *ρ–T* slopes than in oxide perovskites, suggesting a distinct origin rooted in anti-perovskite topology. The coexistence of these two channels explains $Ni_3InN$'s unique transport: robust, high-mobility conduction from boundary states and tunable, strain-sensitive behavior from bulk bands.

**Discussion and Conclusions**

In summary, the stabilization of a rare low-valence state of Ni, pronounced longitudinal rigidity, and a Young's modulus comparable to many functional substrate materials make $Ni_3InN$ as a highly promising electronic and mechanically robust antiperovskite nitride. We have successfully fabricated high-quality single-crystalline $Ni_3InN$ thin films on various substrates and demonstrated the unique strain relaxation mechanisms governed by the material's exceptional stiffness and intrinsic orientation preferences, which sharply differ from conventional perovskite oxides. Atomic-resolution analysis not only confirms the high crystalline quality and sharp interfaces, but also reveals microscopic interfacial configurations such as strain-induced lattice distortions and atomically defined bonding at the heterointerfaces. This fundamental distinction highlights that, whereas oxides are suited for smooth strain engineering, $Ni_3InN$-based films offer a pathway for strain-mediated control of crystalline orientation, thereby opening opportunities for designing antiperovskite heterostructures with tunable anisotropic properties. Electrically, the $Ni_3InN$ thin films exhibit exceptional conductivity and maintain a remarkably stable resistivity across an extensive temperature spectrum. This characteristic renders them highly appropriate for applications including high-accuracy resistors, stable sensors, and other high-precision instruments that function within various temperature conditions. Together, these findings establish $Ni_3InN$ thin films as a new experimental platform for probing emergent electronic phenomena at the intersection of perovskite physics and antiperovskite chemistry. By uniting epitaxial stabilization, interface-sensitive growth, and detailed theoretical modeling, our work opens a pathway to discover novel correlated phases in anti-perovskite nitrides and demonstrates their potential as a counterpart to oxide perovskites in the quest for exotic quantum states, paving the way for their integration into next-generation electronic and spintronic devices.

**Materials and Methods**

**Synthesis of high-quality epitaxial nitride thin films**

Polycrystalline $Ni_3InN$ target were synthesized through a high-pressure reaction route using a mixture of $In_2O_3$, Ni and $NaNH_2$ powders. The mixed powder was sintered at 5 GPa and 900 °C for 60 min at the High-Pressure Lab of South University of Science and Technology (SUSTech). Powder X-ray diffraction measurements on the target demonstrate a stoichiometric and correct chemical composition. Single-crystalline antiperovskite nitride $Ni_3InN$ thin films were fabricated by pulsed laser deposition (PLD) using a XeCl excimer laser with 308 nm wavelength. During the thin film deposition, the energy density and laser frequency maintain ~1.6 J/cm$^2$ and 5 Hz. To systematically investigate the effects of strain on $Ni_3InN$ thin films, all films were grown on (001)-oriented $LaAlO_3$, $SrTiO_3$, and (011)-oriented $DyScO_3$ substrates at a temperature of 500 °C and a base pressure of $1 \times 10^{-8}$ Torr. The thickness of $Ni_3InN$ film, approximately 30 nm, was controlled by counting the number of laser pulses and further confirmed by X-ray reflectometry. The samples were cooled down slowly to room temperature at a rate of –5°C under high-vacuum atmosphere (~ $10^{-7}$-$10^{-8}$ Torr), keeping the same as the growth conditions.

**Structural characterization and electronic state measurements**

The thin film quality was examined by high-resolution four-circle X-ray diffractometer (Malvern Panalytical, X'Pert3 MRD) with $\theta$-$2\theta$ line scans and rocking curve scans. Phi scans and off-specular RSM are conducted to determine the epitaxial growth and in-plane strain states of $Ni_3InN$ thin films. The temperature-dependent X-ray diffraction (XRD) was measured using a Bruker D8 Discovery diffractometer. X-ray reflectivity (XRR) measurements were performed to check the layer thickness and interface roughness. Electrical measurements including temperature-dependent resistivity, magnetoresistance (MR) and Hall resistance were conducted in a Physical Property Measurement System (PPMS), using standard van der Pauw method. The optical conductivity of different strained RP nickelate thin films were measured at room temperature using a commercial optical ellipsometer (J. A. Woollam Co., Inc.).

**STEM characterizations and analysis**

Cross-sectional transmission electron microscopy (TEM) samples were fabricated by standard ion beam (FIB) lift-off-process and measured in scanning mode using a JEM ARM 200CF microscope at the Institute of Physics, Chinese Academy of Sciences. The high-angle annular dark field (HAADF) images were collected along [100] and [110] orientation for confirming the crystal structure. All scanning transmission electron microscopy (STEM) images were analyzed using Gatan DigitalMicrograph. Electron-energy-loss-spectroscopy (EELS) mappings were performed at Ni *L*-, In *L*-, La *M*-, Al *K*-, Sr *L*-and Ti *L*-edges to evaluate the interfacial mixing.

**First-principles calculations**

Density-Functional Theory (DFT) (*57,58*) calculations were performed to investigate the structural relaxation, electronic structure, and elastic properties of $X_3InN$ materials. The simulations were conducted using the Vienna *Ab initio* Simulation Package (VASP) (*59,60*) and WIEN2k (*61,62*), employing the Perdew–Burke–Ernzerhof (PBE) form of the Generalized Gradient Approximation (GGA) for the exchange-correlation functional (*63*). A plane-wave energy cutoff of 500 eV and a Monkhorst-Pack *k*-point mesh of 13×13×13 was used for Brillouin zone integration. Phonon spectra were computed using both the frozen-phonon (finite displacement) method (*64*) and density-functional perturbation theory (DFPT) (*65*), implemented via the Phonopy package (*66*) interfaced with VASP, based on the relaxed ground-state structures of each $X_3InN$ compound.

**Acknowledgements**

We thank Ling Lu's group at IOP-CAS for helping the optical conductivity measurements and Shuai Dong's group at Southeast University for general theoretical inputs. **Funding:** This work was supported by the Beijing Natural Science Foundation (Grant No. JQ24002 to E.J.G. and IS25040 to S.Choi), the National Natural Science Foundation of China (Grant Nos. 12422407 to L.S., U22A20263 and 52250308 to E.J.G., 12304158 to Q.J., and 12474096 to C.W.), the National Key Basic Research Program of China (Grant Nos. 2020YFA0309100 to E.J.G.), the CAS Project for Young Scientists in Basic Research (Grant No. YSBR-084 to E.J.G.), the Guangdong Basic and Applied Basic Research Foundation (Grant No. 2022B1515120014 to E.J.G.), the Guangdong-Hong Kong-Macao Joint Laboratory for Neutron Scattering Science and Technology, the China Postdoctoral Science Foundation (Grant No. 2022M723353 to E.J.G.), and the International Young Scientist Fellowship of IOP-CAS to S.Choi. Theoretical calculations have been mainly done in Northwest University (Xi'an). **Author Contributions:** E.J.G. initiated the research and supervised the project. The $Ni_3InN$ target was prepared by S.W. at SUSTech University. The sample growth and processing were carried out by T.C. with help from Q.W., D.R., and S. Choi. TEM lamellae were prepared via FIB milling, and TEM experiments were conducted by Q.H.Z. Transport measurements were carried out by T.C. with assistance from A.X. and H.J. DFT calculations were performed by Z.X., X. D. Z. under the supervision of L.S. C.G., C.W., and K.J.J. contributed to discussions and provided valuable suggestions during manuscript preparation. T.C., Z.X., L.S., and E.J.G. wrote the manuscript. All authors have discussed the results and the interpretations. **Competing interests:** The authors declare that they have no competing interests. **Data and Materials availability:** All data needed to evaluate the conclusions in the paper are present in the paper and/or the Supplementary Materials. Additional datasets generated during and/or analyses during the current study are available from the first author (T.C.) and corresponding authors (L.S., E.J.G., and K.J.J.) on reasonable request.

**Figures and Tables**

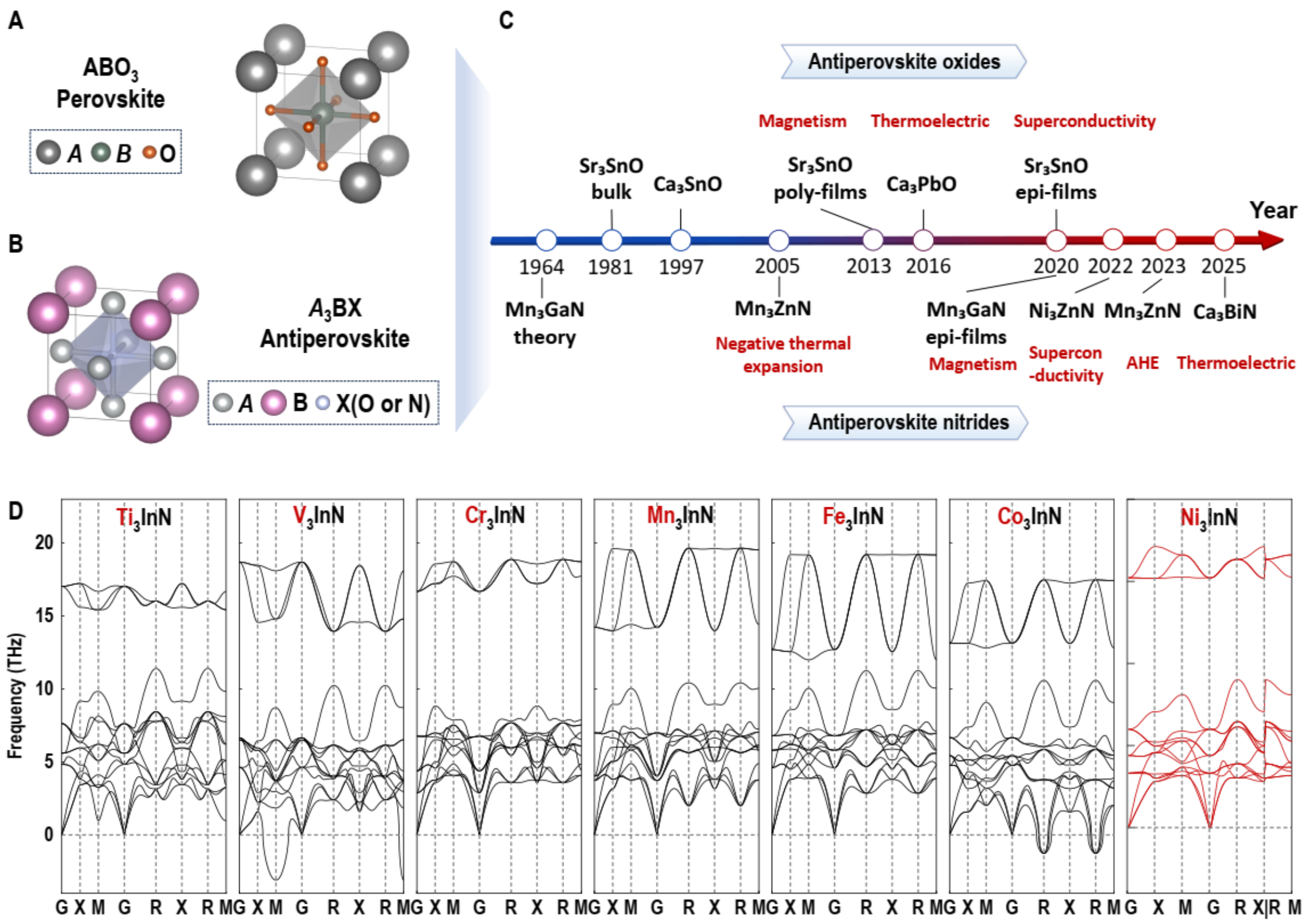

**Fig. 1. Evolution of antiperovskite-type materials.** Schematic illustrations of the crystal structures of (**A**) perovskite ($ABX_3$, where X = O or N) and (**B**) antiperovskite ($A_3BX$, where X = O or N) oxides/nitrides. (**C**) Timeline of the discovery of key physical properties in antiperovskite oxides and nitrides. (**D**) DFT phonon spectra for $A_3InN$ antiperovskite nitrides, including $Ti_3InN$, $V_3InN$, $Cr_3InN$, $Mn_3InN$, $Fe_3InN$, $Co_3InN$ and $Ni_3InN$, where $V_3InN$ and $Co_3InN$ are dynamically unstable.

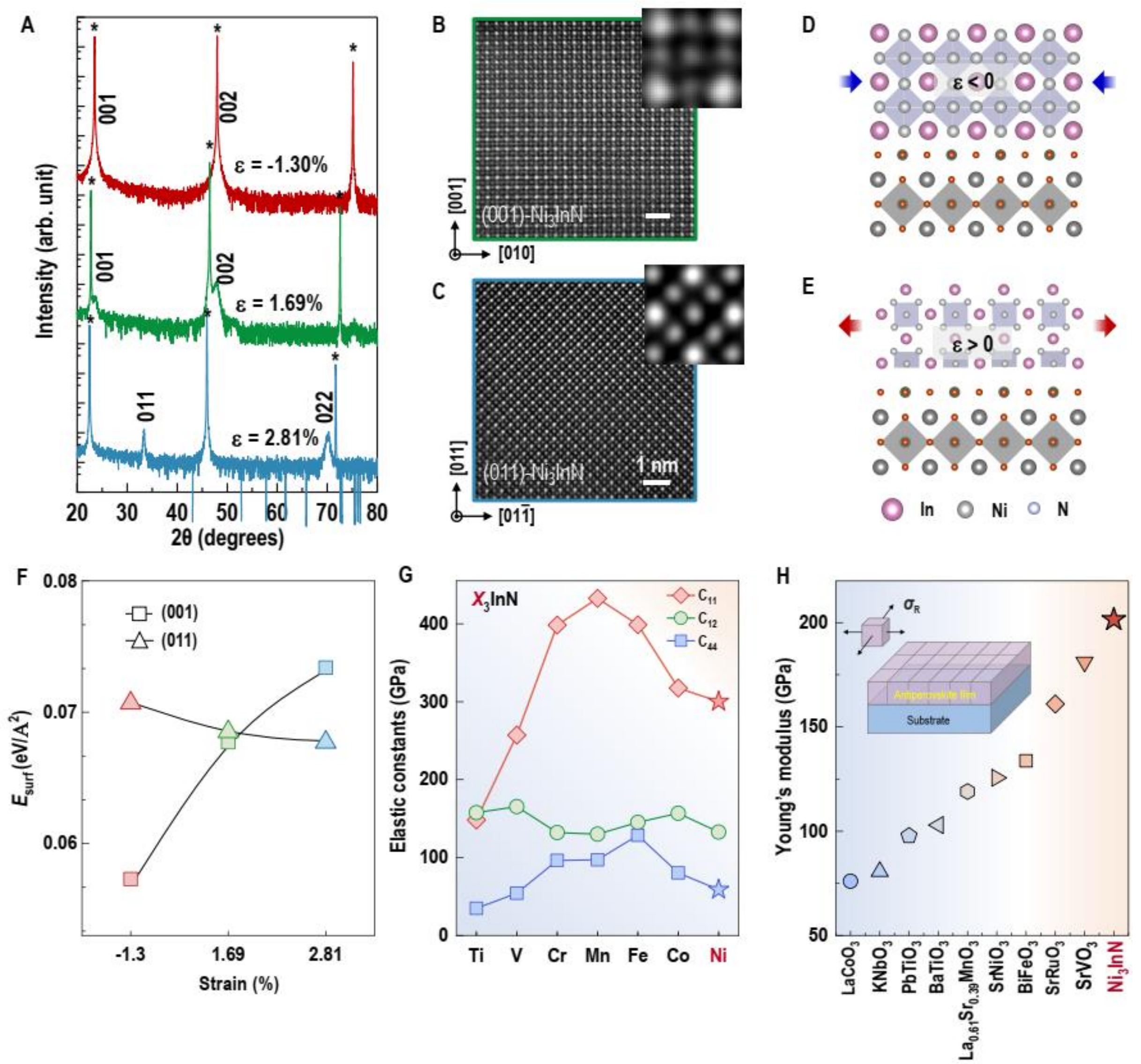

**Fig. 2. Structural characterization of $Ni_3InN$ thin films.** (**A**) X-ray diffraction *θ*-2*θ* scans of $Ni_3InN$ films grown on $LaAlO_3$ ($\varepsilon$ = -1.30%), $SrTiO_3$ ($\varepsilon$ = 1.69%), and $DyScO_3$ ($\varepsilon$ = 2.81%) substrates (indicated with "*"). (**B**) High-resolution HAADF-STEM image from a (001)-oriented $Ni_3InN$ film grown on $SrTiO_3$. Inside shows a representative unit cell. (**C**) HAADF-STEM image of (011)-oriented $Ni_3InN$ film on $DyScO_3$, inset showing a magnified unit cell. (**D**) and (**E**) Schematic illustrations
(**F**) Formation energy $E_{surf}$ of (001)- and (011)-oriented $Ni_3InN$ under different strain states. (**G**) Calculated elastic constants $C_{11}$, $C_{12}$, and $C_{44}$ of $X_3InN$ as a function of X-site composition (X= Ti, V, Cr, Mn, Fe, Co, Ni). $Ni_3InN$ exhibits higher $C_{11}$ and lower $C_{44}$, pointing to enhanced longitudinal stiffness but reduced shear rigidity. (**H**) Comparison of Young's modulus of $Ni_3InN$ with the previously reported conventional perovskite oxides, indicating that $Ni_3InN$ possesses enhanced mechanical rigidity comparable to the extensively studied perovskite oxides. The inset in (**H**) illustrates a schematic diagram of the equibiaxial tensile strain applied to the film by the substrate.

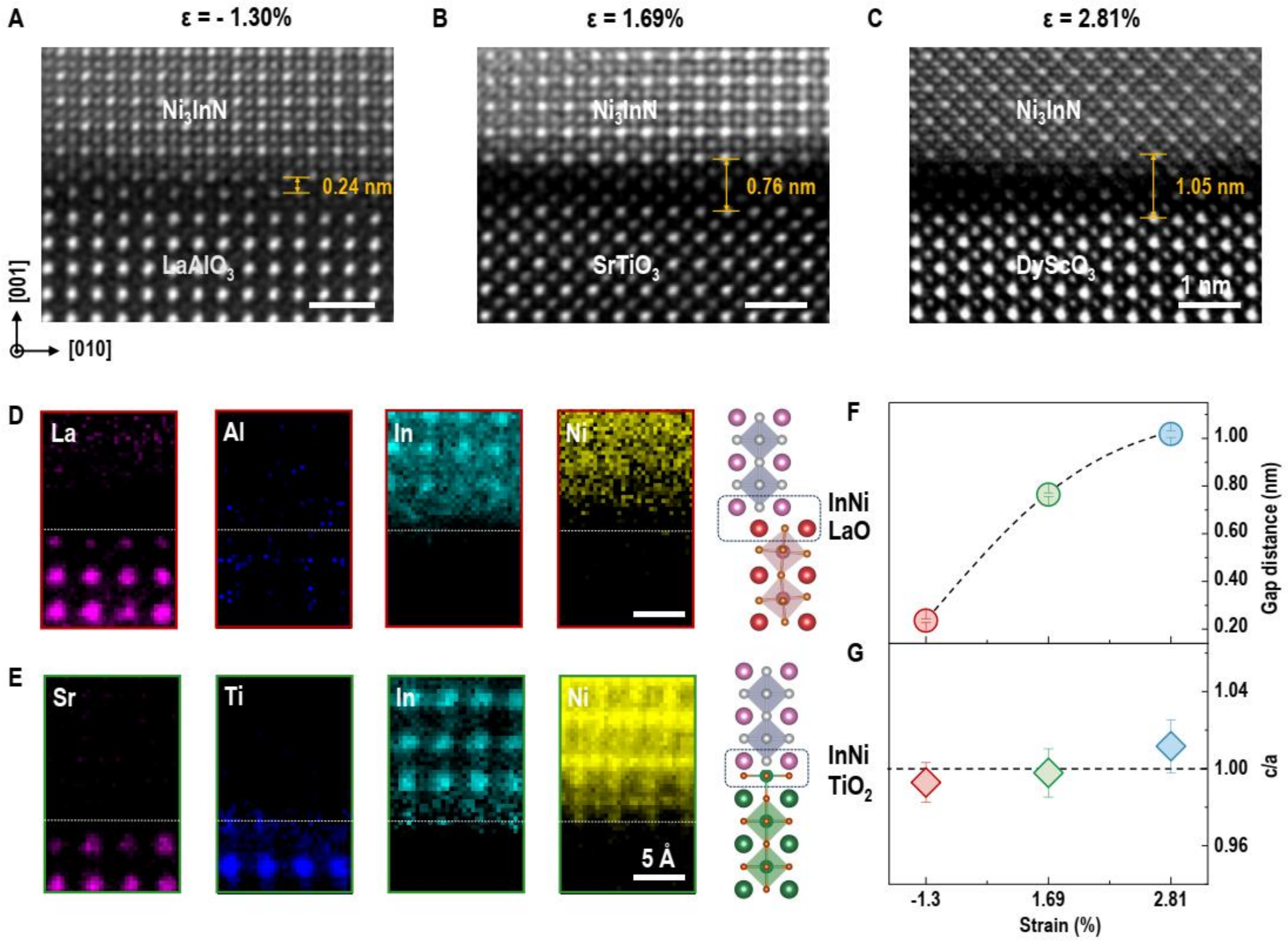

**Fig. 3. Atomic-scale characterization of $Ni_3InN$ interfacial structures on three distinct substrates.** (**A-C**) HAADF-STEM images of the $Ni_3InN$ thin films grown on $LaAlO_3$, $SrTiO_3$ and $DyScO_3$ substrates taken along the [100] zone axis. The interfacial regions with variable gap distances are annotated in yellow. (**D**) Compositional EELS mapping of $Ni_3InN/LaAlO_3$ taken simultaneously at La *M*-, Al *K*-, In *M*-, and Ni *L*-edges, respectively. Representation of the $Ni_3InN/LaAlO_3$ heterointerface as a stacking of atomic unit cell planes, showing a half-unit-cell lateral lattice misalignment between adjacent atomic planes. (**E**) Atomic-scale EELS mapping of $Ni_3InN/SrTiO_3$ taken simultaneously at Sr *L*-, Ti *L*-, In *M*-, and Ni *L*-edges, respectively. $Ni_3InN$ is epitaxially grown on $SrTiO_3$ via cube-on-cube orientation. (**F**) Gap distances between $Ni_3InN$ thin films and substrates as functions of strain. (**G**) Strain-dependent *c*/*a* ratio of $Ni_3InN$ thin films.

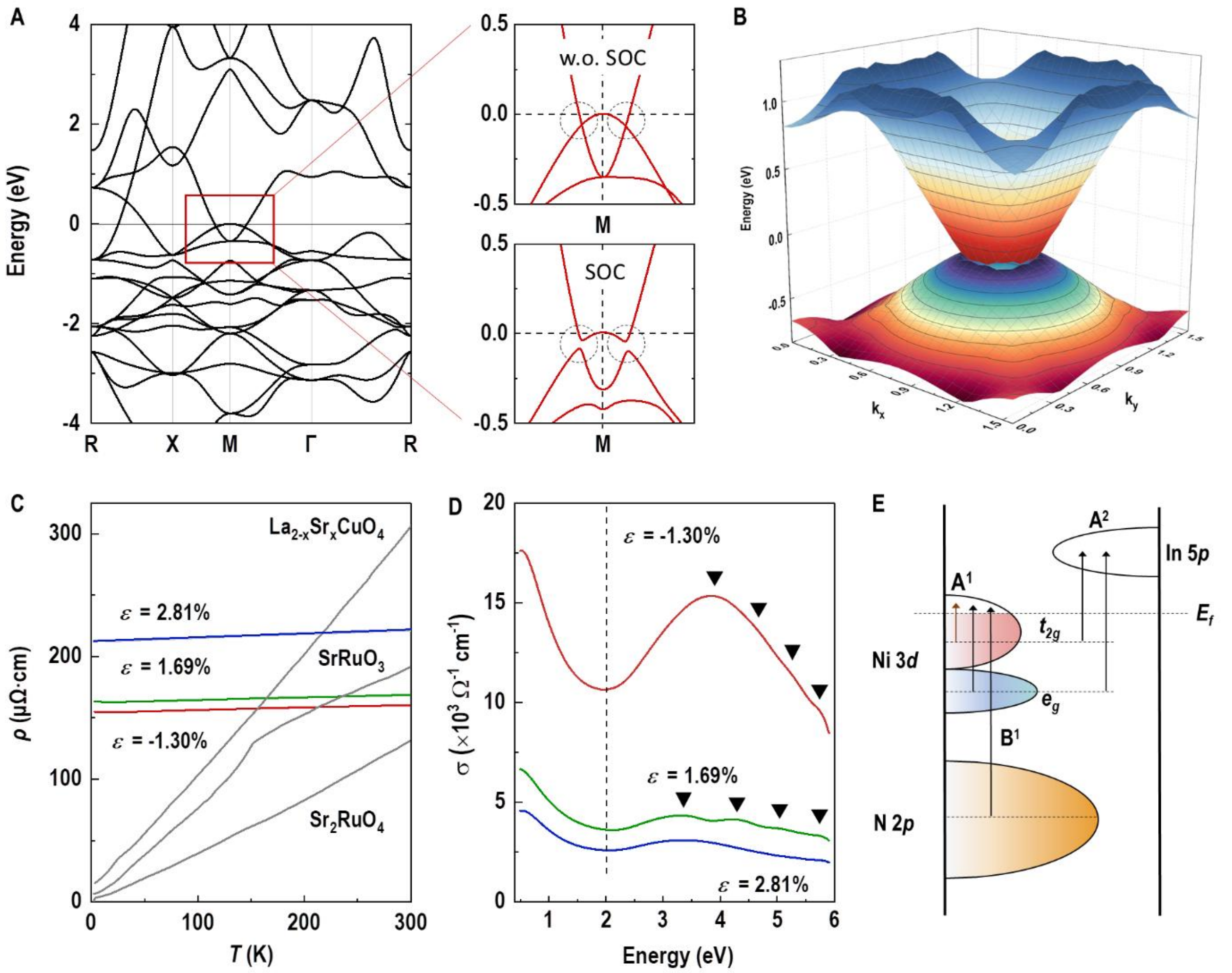

**Fig. 4. Band structure calculations and electronic properties of $Ni_3InN$.** (**A**) DFT band structure of $Ni_3InN$ highlighting empty In-*p* states and fully filled N-*p* bands, consistent with an $In^{1+}/N^{3-}$ charge state and an average Ni valence of $Ni^{+2/3}$. The right side presents the band structures around M point without (w.o.) spin-orbital coupling (SOC) and with SOC. (**B**) Three-dimensional band structure of $Ni_3InN$. (**C**) Temperature-dependent resistivity ($\rho$) of $Ni_3InN$ thin films on various substrates. The temperature-dependent $\rho$ of $La_{2-x}Sr_xCuO_4$, $SrRuO_3$, and $Sr_2RuO_4$ were shown for comparison. (**D**) Optical conductivities of $Ni_3InN$ thin films on various substrates, obtained from spectroscopic ellipsometry at room temperature. The arrows indicate the optical excitation peaks. (**E**) Schematic of band structure, extracted from optical conductivity measurements.

**Table 1. Summary of Young modulus for conventional perovskite oxides and antiperovskite nitride $Ni_3InN$ presented in this work.**

| Materials | Young modulus/GPa | Refs. |
|---|---|---|
| $LaCoO_3$ | 76 | (*37*) |
| $KNbO_3$ | 80.64 | (*41*) |
| $PbTiO_3$ | 97.75 | (*41*) |
| $BaTiO_3$ | 102.9 | (*41,43*) |
| $La_{0.61}Sr_{0.39}MnO_3$ | 120 | (*38,39*) |
| $BiFeO_3$ | 133.73 | (*42*) |
| $SrRuO_3$ | 161 | (*40*) |
| $SrVO_3$ | 181.1 | (*36*) |
| $Ni_3InN$ | 201.5 | / |

# Supplementary Materials for

## Unusual strain relaxation and Dirac semimetallic behavior in epitaxial antiperovskite nitrides

Ting Cui *et al.*
*Corresponding author. Email: kjjin@iphy.ac.cn, siliang@nwu.edu.cn, and ejguo@iphy.ac.cn

**This PDF file includes:**

figs. S1 to S14

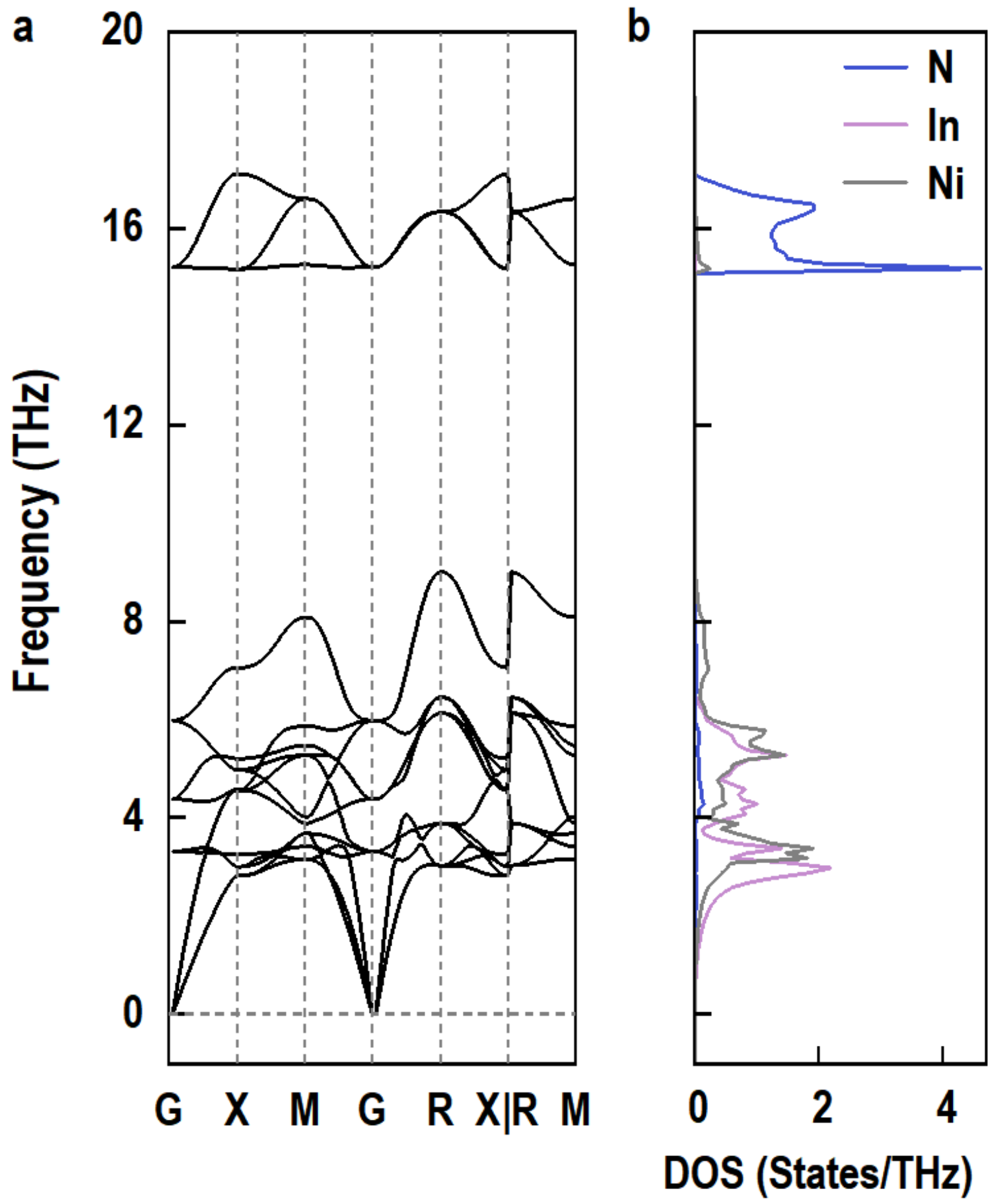

**fig. S1. Phonon characterization of $Ni_3InN$.** (a) Calculated phonon dispersion and (b) density of states of $Ni_3InN$ showing no imaginary frequencies, indicating dynamical stability.

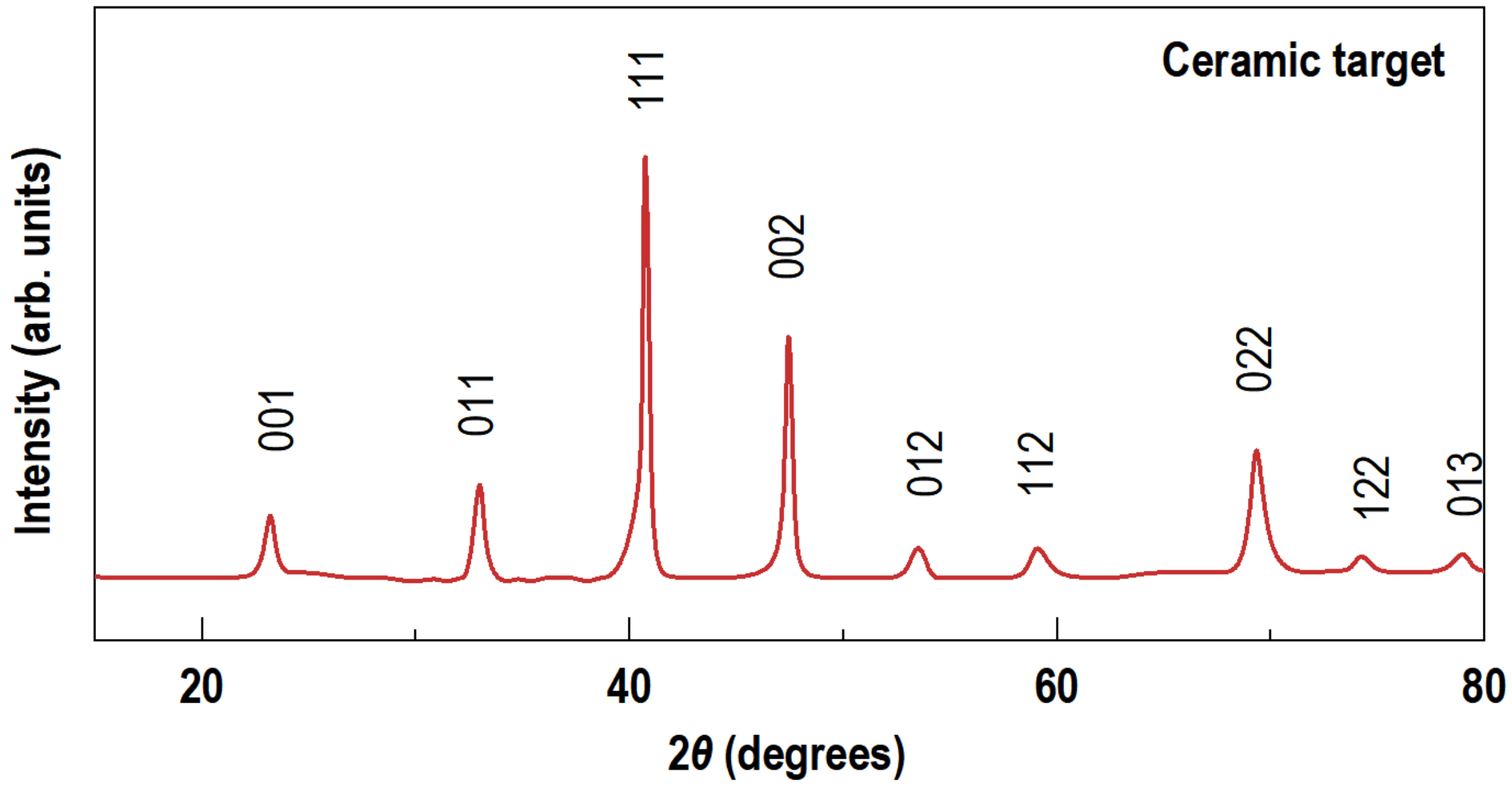

**fig. S2. Powder XRD $\theta$-2$\theta$ scan of $Ni_3InN$ ceramic target.**

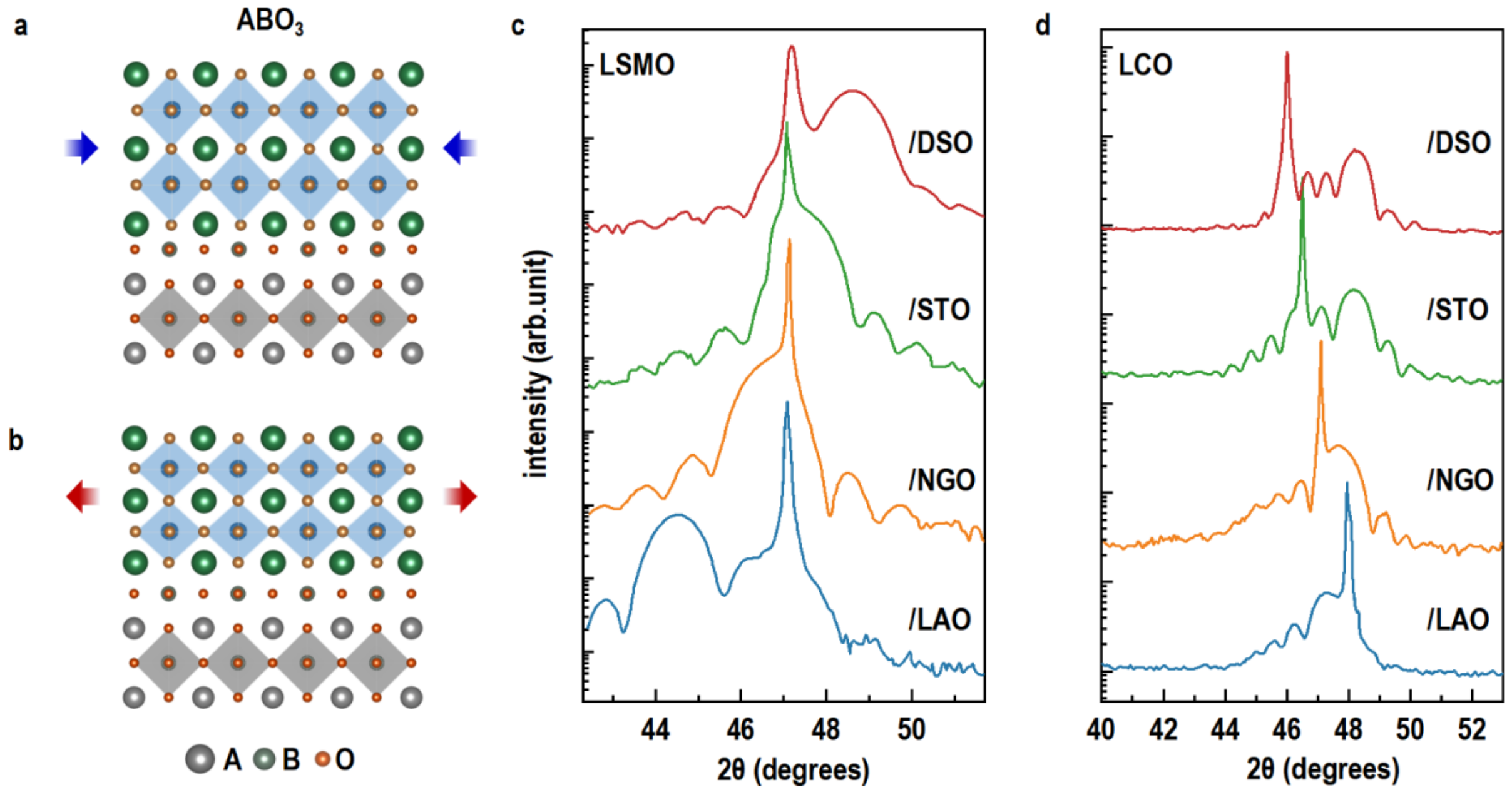

**fig. S3. Strain effects on conventional perovskite oxides.** (a) and (b) Schematic illustrations of lattice deformation in perovskite thin films under compressive and tensile strain, respectively, induced by the substrate's lattice misfit. (c) and (d) XRD *θ-2θ* scans of $La_{0.7}Sr_{0.3}MnO_3$(LSMO) and $LaCoO_3$(LCO) thin films grown on substrates [$DyScO_3$(DSO), $SrTiO_3$(STO), $NdGaO_3$(NGO), and $LaAlO_3$(LAO)] with varying lattice constants. The films are epitaxially and coherently grown, maintaining the in-plane lattice constants of the substrates. The out-of-plane lattice constants of the films vary systematically in response to the strain, while the unit cell volume remains constant.

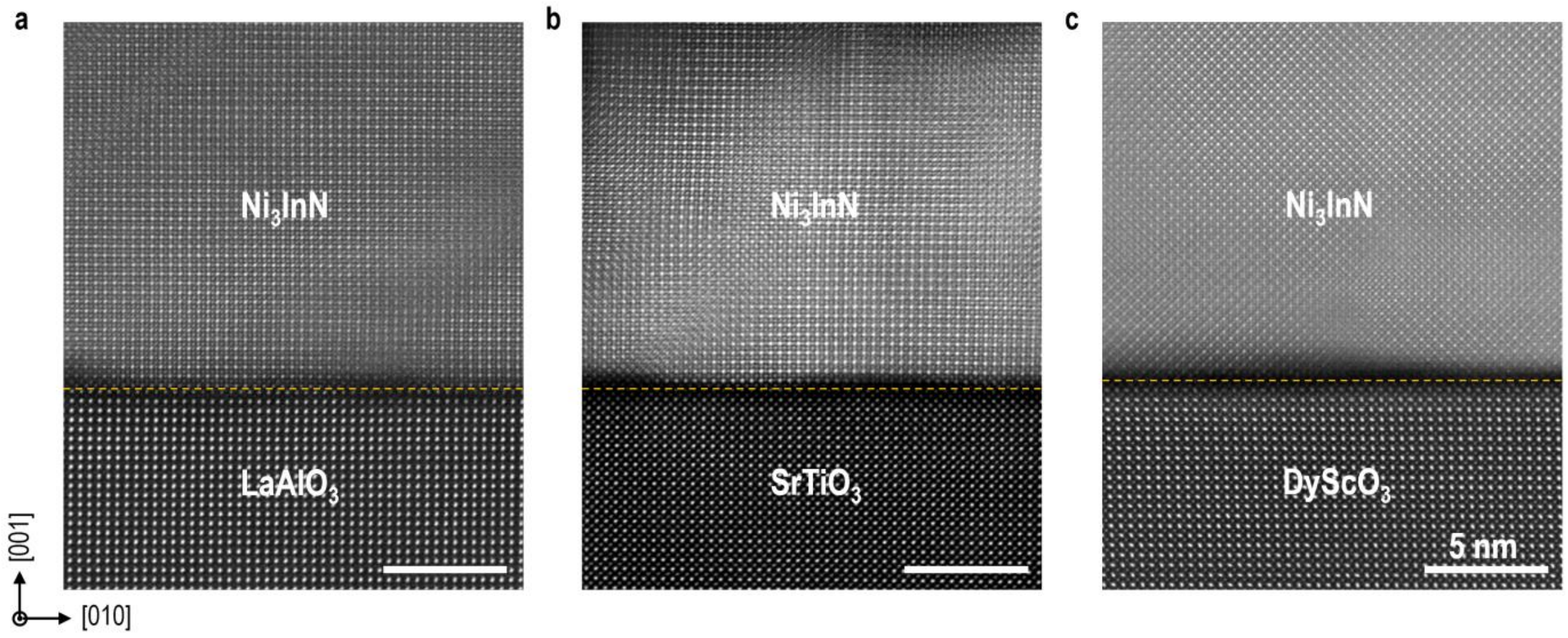

**fig. S4. STEM-HAADF images of $Ni_3InN$ thin films on various substrates.** (a)-(c) STEM-HAADF images of $Ni_3InN$ thin films grown on $LaAlO_3$, $SrTiO_3$, and $DyScO_3$ substrates, respectively. Yellow dashed lines highlight the interfaces between the $Ni_3InN$ films and the substrates. The films exhibit high crystallinity, with the interfacial gap distances increasing progressively from the compressive-strain region to the tensile-strain region.

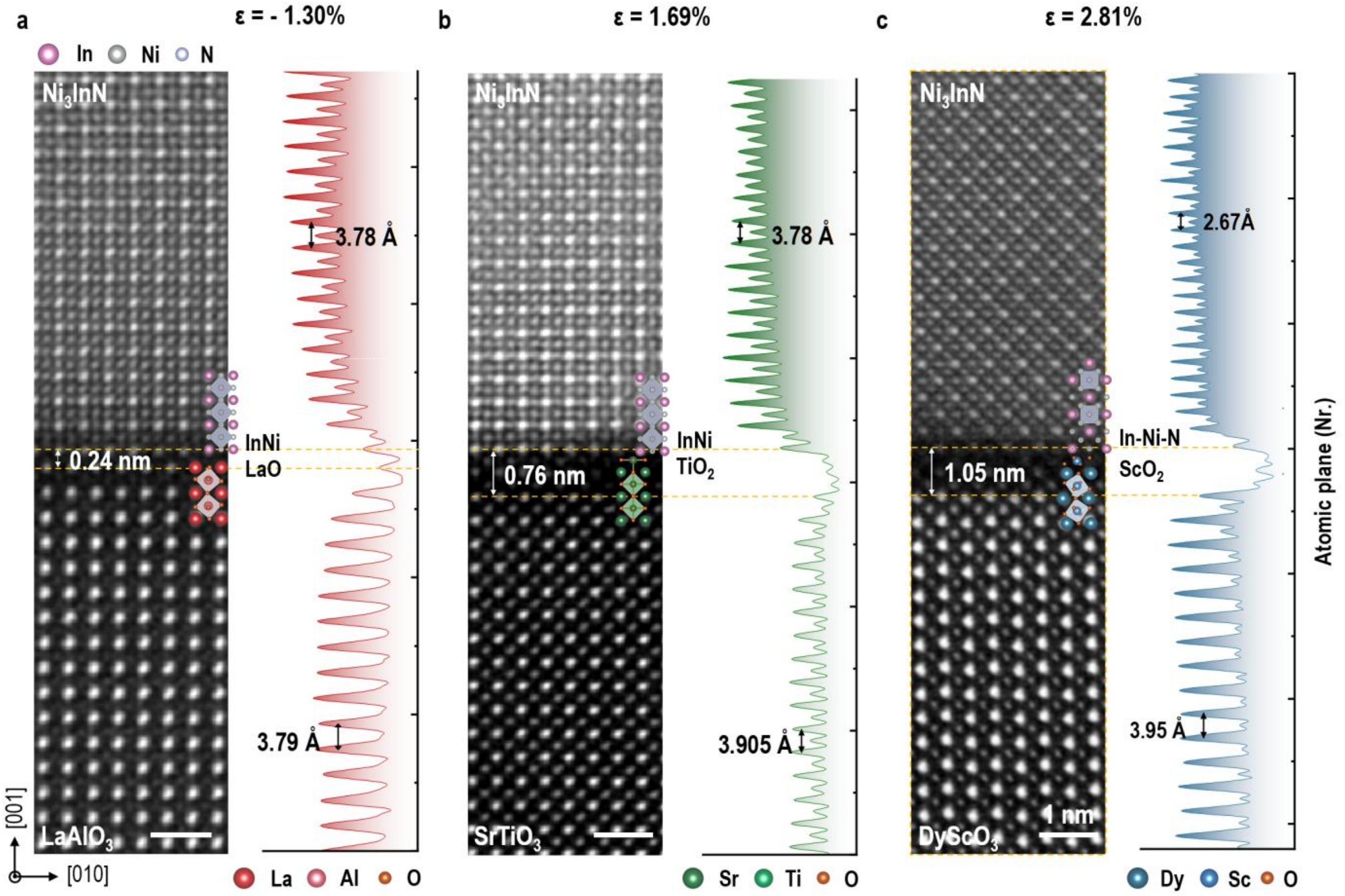

**fig. S5. Microstructural characterization of $Ni_3InN$ thin films.** Cross-sectional HAADF-STEM images and corresponding layer-resolved intensity profiles of (a) $Ni_3InN/LaAlO_3$, (b) $Ni_3InN/SrTiO_3$, and (c) $Ni_3InN/DyScO_3$ interfaces. The atomic configurations at the heterointerfaces are shown, with interfacial gaps (highlighted in yellow dashed lines) labeled for each system. $Ni_3InN$ thin films grow along the [001] orientation on $LaAlO_3$ (ε = -1.30%) and $SrTiO_3$ (ε = 1.69%) substrates, exhibiting small compressive and tensile strains, respectively. In contrast, films grown on the large tensile-strain $DyScO_3$ substrate (ε = 2.81%) exhibit [011]-oriented growth.

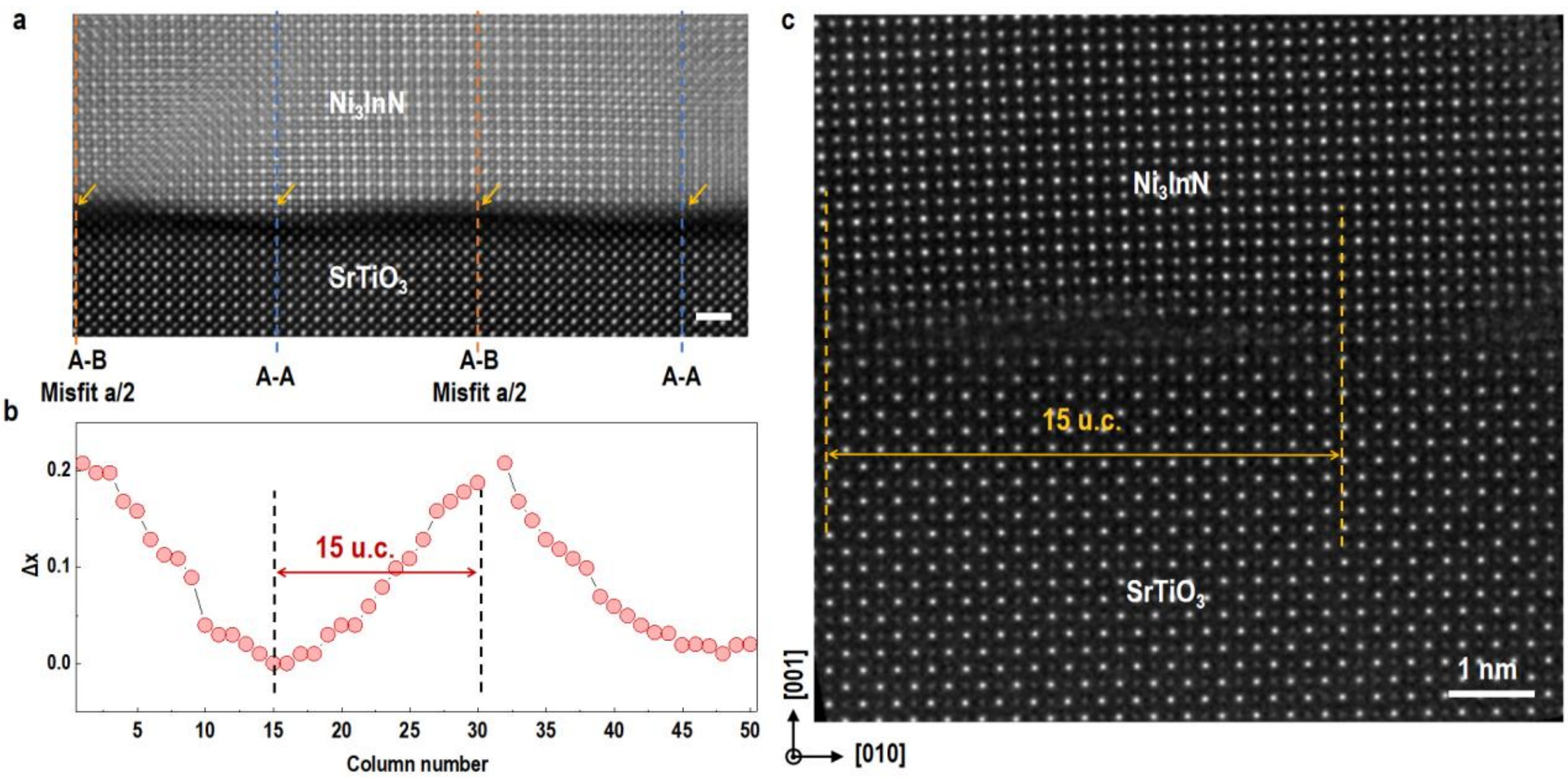

**fig. S6. Atomic-scale characterization of $Ni_3InN/SrTiO_3$ interface.** (a) A representative HAADF-STEM image of the $Ni_3InN/SrTiO_3$ interface. (b) Displacement distances as a function of lateral atomic column number. The lattices of the $Ni_3InN$ film and $SrTiO_3$ substrate realign periodically every ~30 unit cells (u.c.), with a lateral displacement of 0.5*a* per 15 u.c. (highlighted with yellow arrows). (c) A zoomed-in 4D HAADF-STEM image of the $Ni_3InN/SrTiO_3$ interface, showing significant atomic displacements and defect formation at the sites of interfacial misfit dislocations.

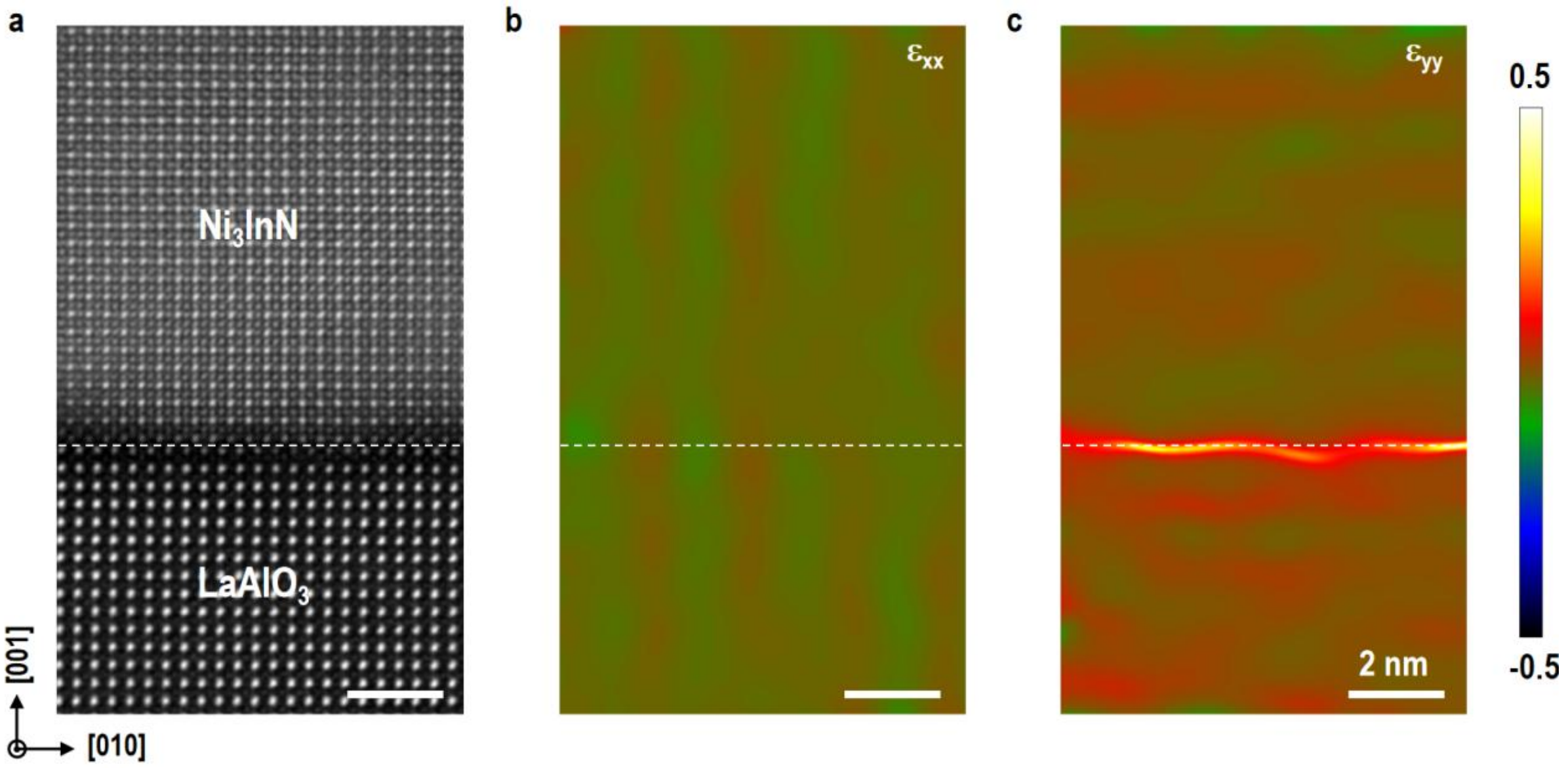

**fig. S7. Strain analysis of $Ni_3InN$ thin films grown on $LaAlO_3$ substrates.** (a) High-resolution HAADF-STEM image of the $Ni_3InN/LaAlO_3$ interface. (b) In-plane strain ($\varepsilon_{xx}$) and (c) out-of-plane strain ($\varepsilon_{\gamma\gamma}$) distributions in the interfacial region.

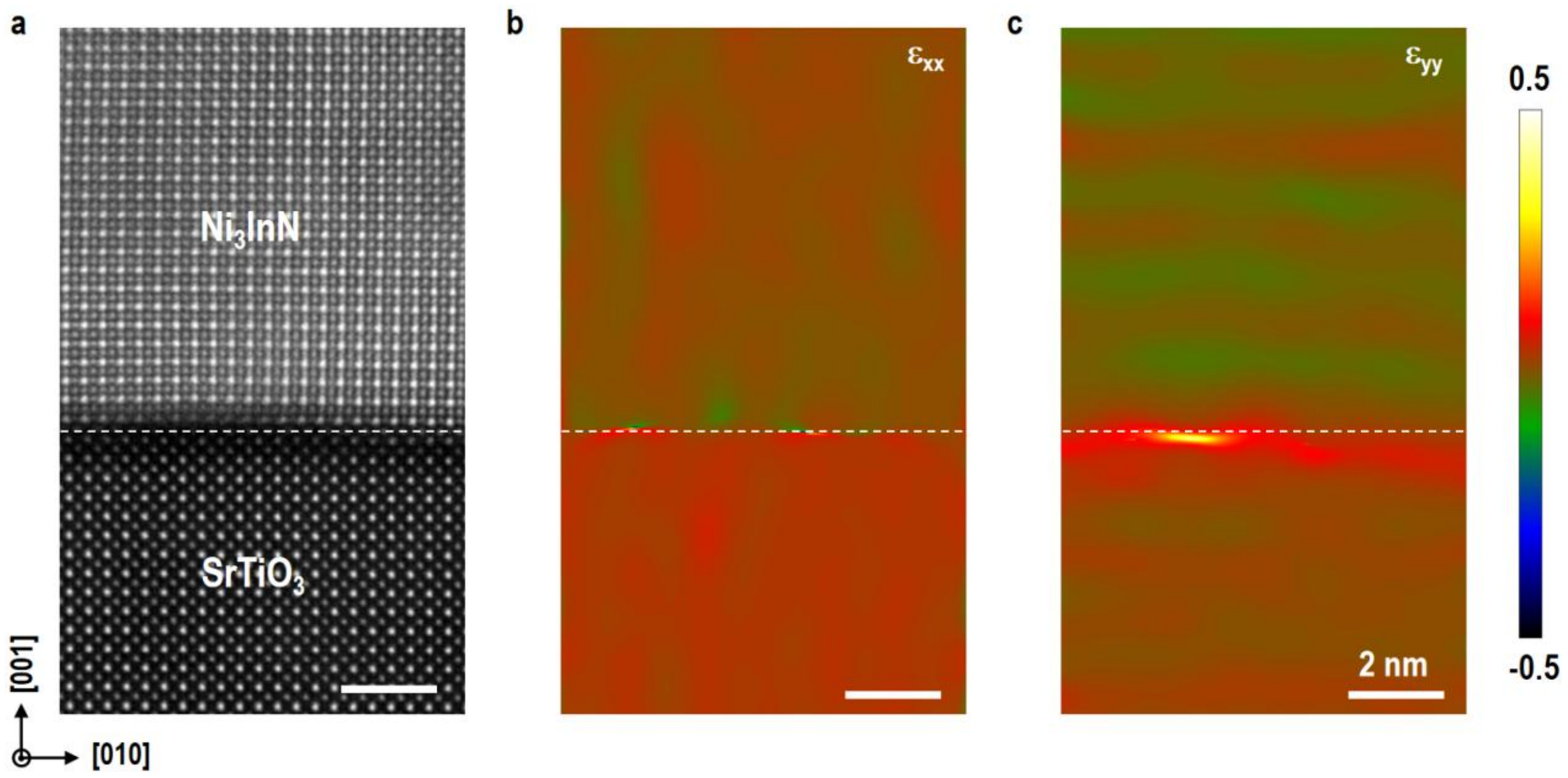

**fig. S8. Strain analysis of $Ni_3InN$ thin films grown on $SrTiO_3$ substrates.** (a) High-resolution HAADF-STEM image of the $Ni_3InN/SrTiO_3$ interface. (b) In-plane strain ($\varepsilon_{xx}$) and (c) out-of-plane strain ($\varepsilon_{yy}$) distributions in the interfacial region.

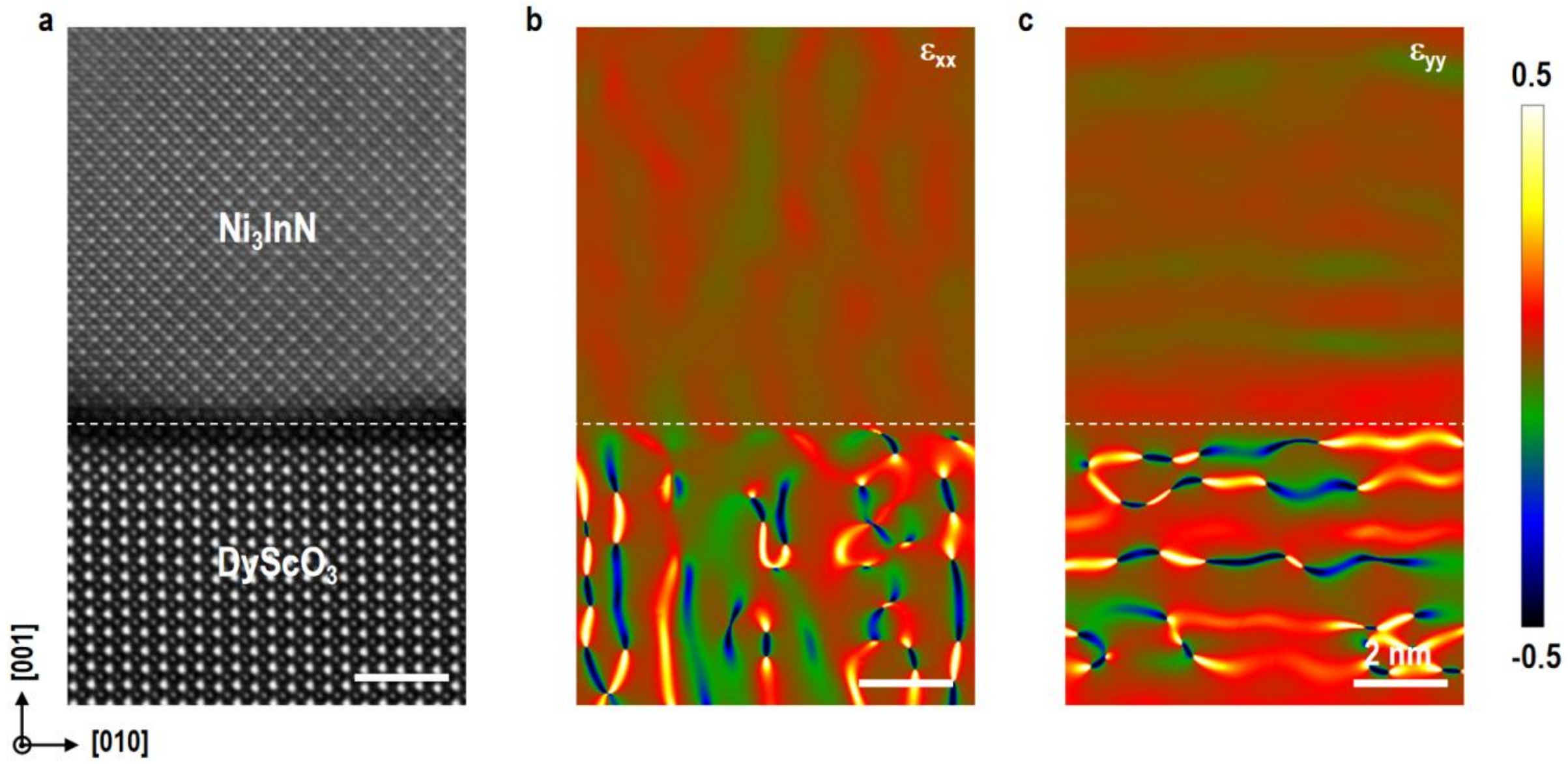

**fig. S9. Strain analysis of $Ni_3InN$ thin films grown on $DyScO_3$ substrates.** (a) High resolution HAADF-STEM image of the $Ni_3InN/DyScO_3$ interface (b) In-plane strain ($\varepsilon_{xx}$) and (c) out-of-plane strain ($\varepsilon_{yy}$) distributions in the interfacial region.

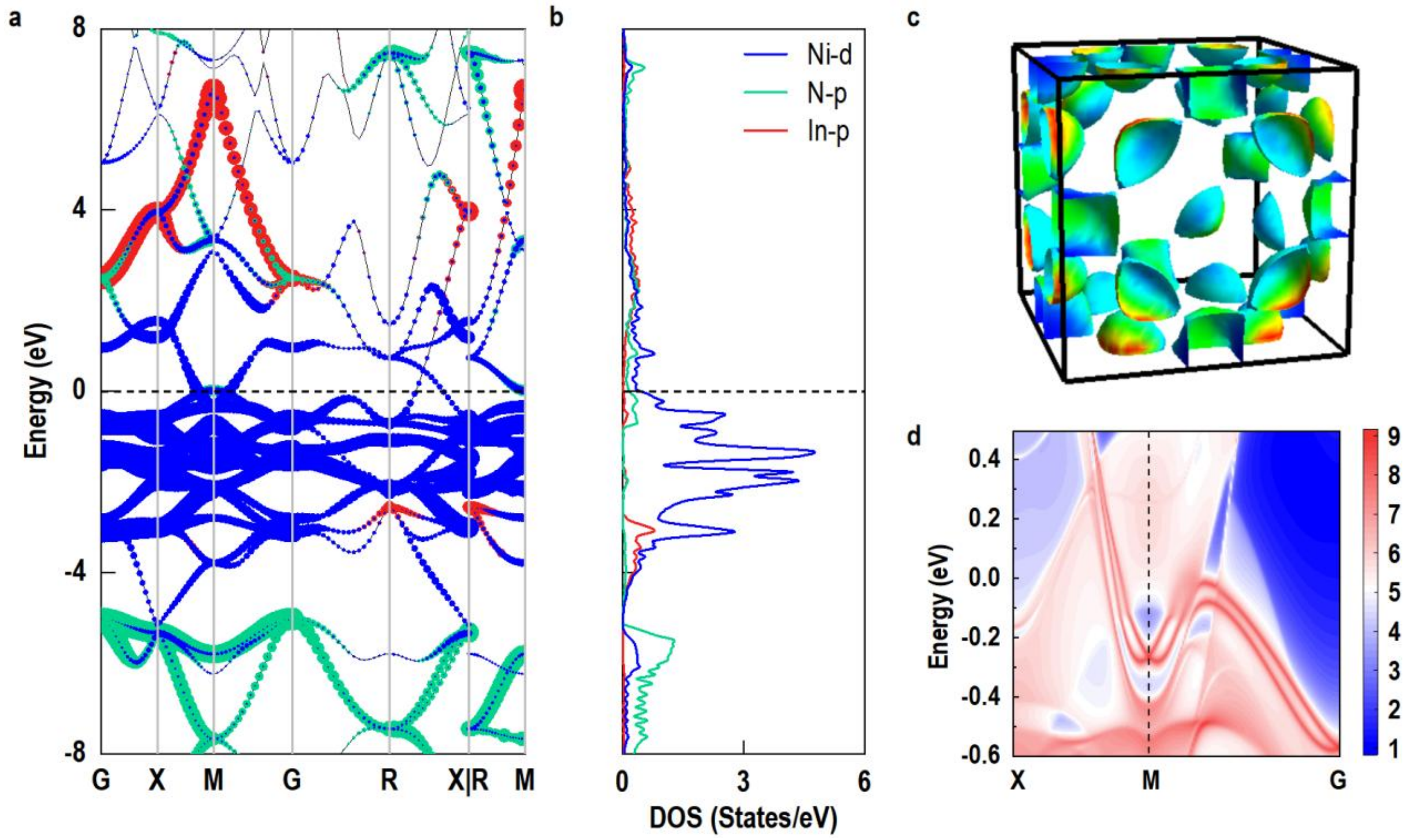

**fig. S10. Band structure calculations of $Ni_3InN$.** (a) DFT band structure, (b) partial density of states, (c) Fermi surface and (d) edge state of $Ni_3InN$ highlighting empty In-p states and fully filled N-p bands, consistent with an $In^{1+}/N^{3-}$ charge state and an average Ni valence of $Ni^{+2/3}$.

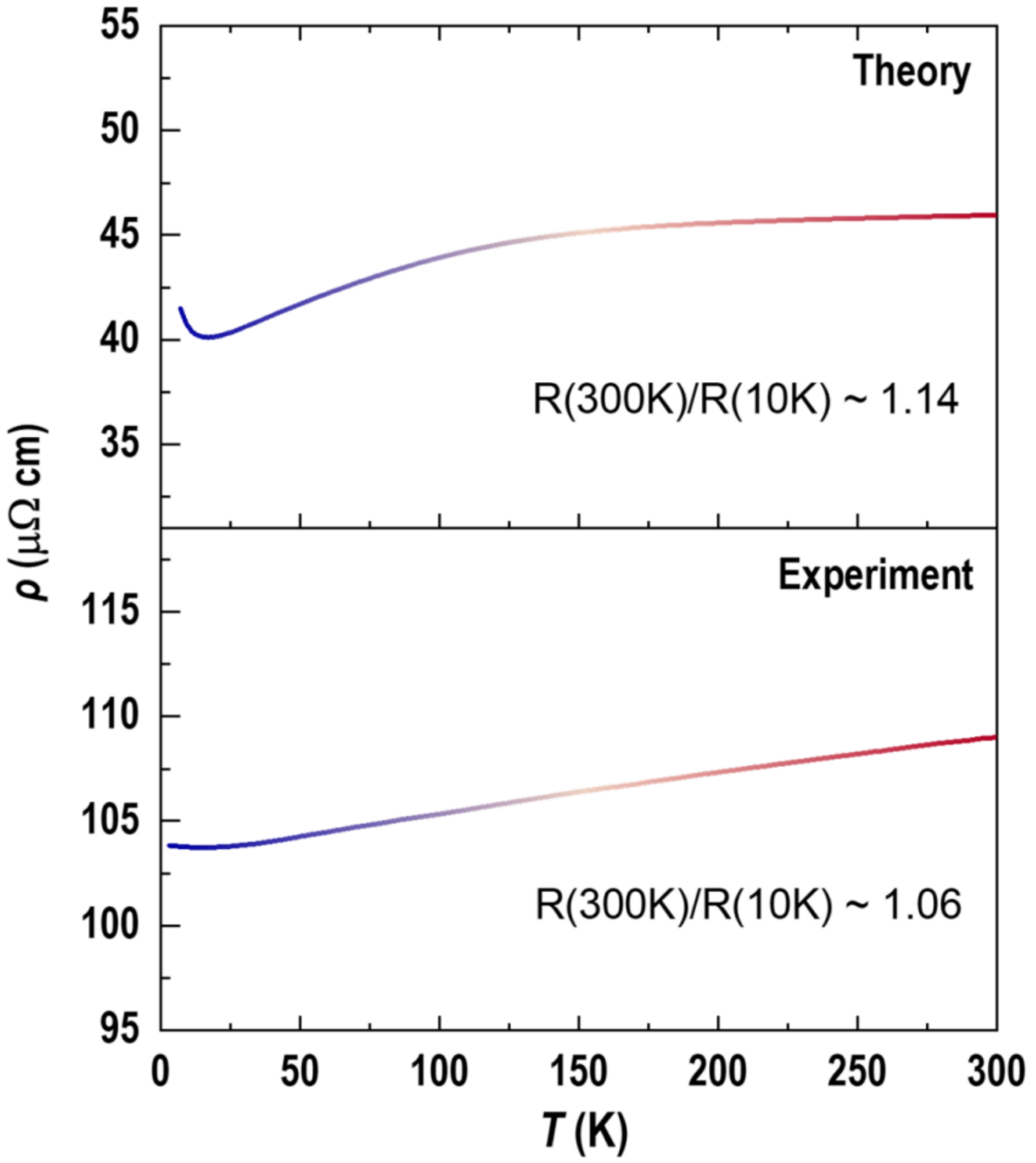

**fig. S11. Comparison of theoretical calculations and experimental results for the transport behavior of antiperovskite $Ni_3InN$ thin films.** Direct comparison of theoretical and experimental results shows similar variation trends in transport behavior, validating the consistency between the two approaches; the resistance ratios at 300 K and 10 K are 1.14 (theoretical) and 1.06 (experimental), respectively, indicating a minor discrepancy and confirming the high accuracy of the theoretical model.

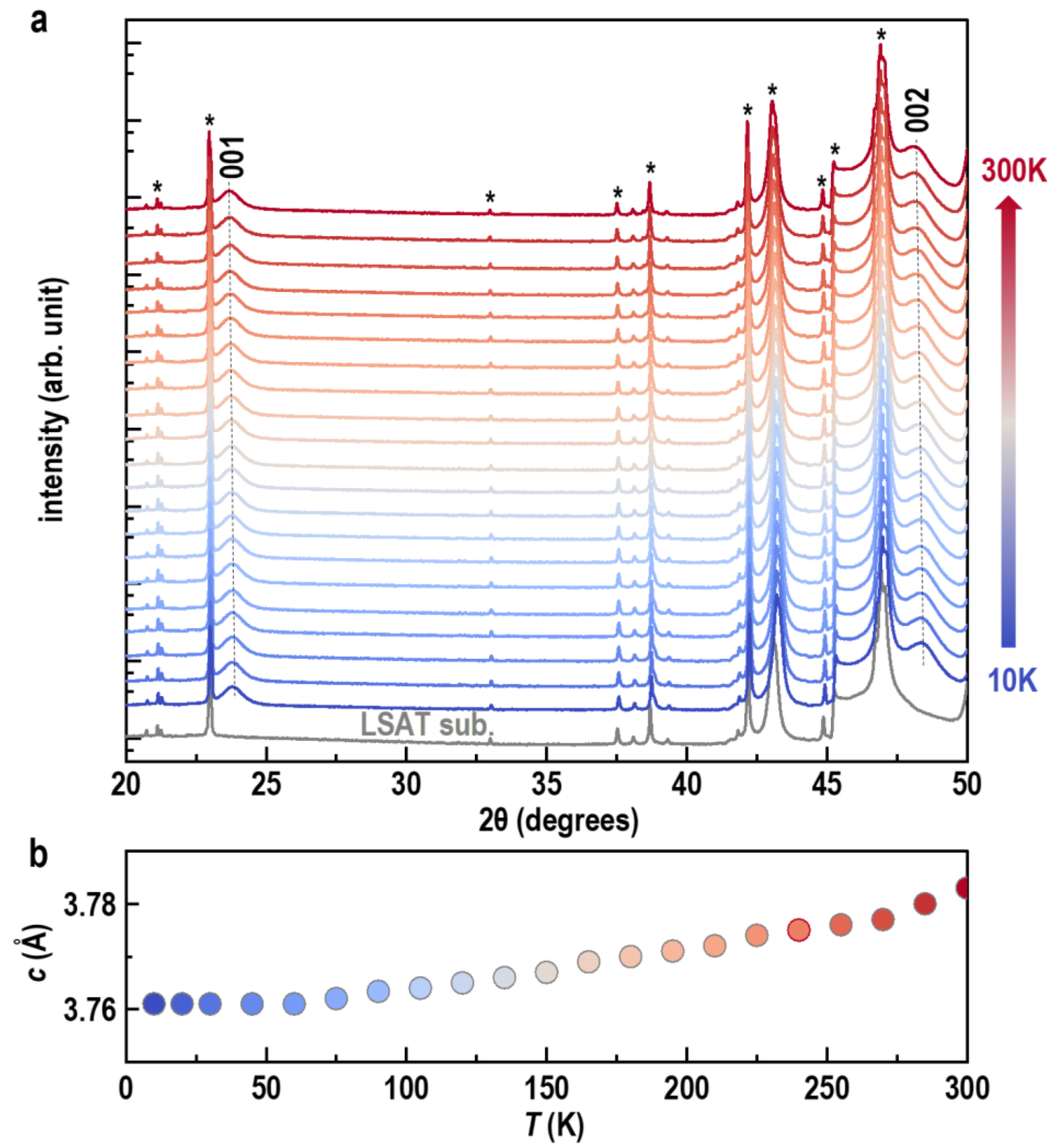

**fig. S12. Lattice stability of $Ni_3InN$ thin films.** (a) Temperature-dependent XRD *θ-2θ* scans showing the 001 and 002 diffraction peaks of $Ni_3InN$, which remain nearly unchanged across the entire temperature range, indicating excellent structural stability. (b) Out-of-plane lattice constant (*c*) of $Ni_3InN$ thin films as a function of temperature, showing a progressive increase due to thermal expansion effects.

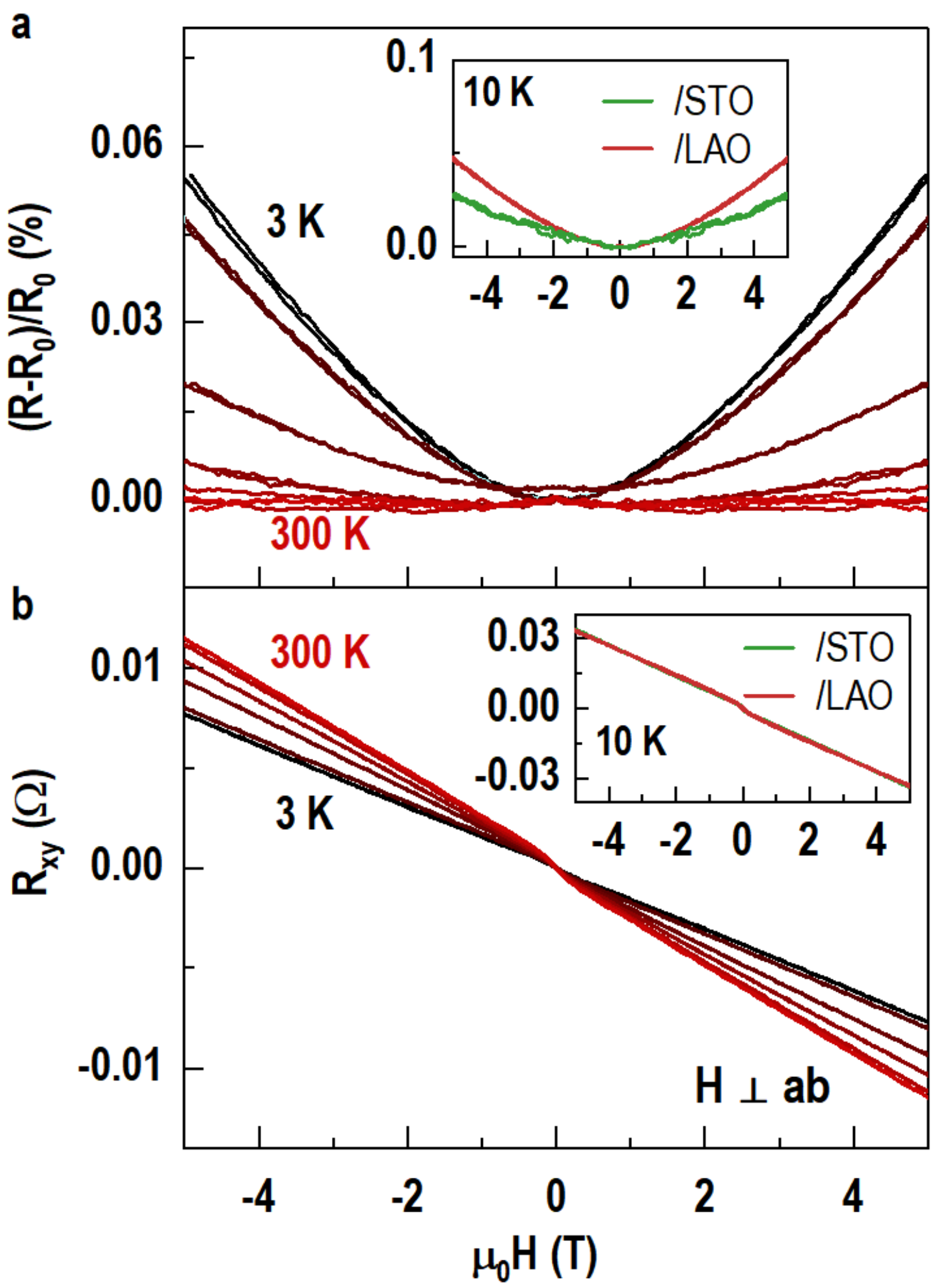

**fig. S13. Magnetotransport properties of $Ni_3InN$ thin films.** (a) Field-dependent magnetoresistance measurements at various temperatures for $Ni_3InN$ thin films grown on $SrTiO_3$ and $LaAlO_3$ substrates. (b) Hall resistance ($R_{xy}$) as a function of magnetic field at different temperatures.

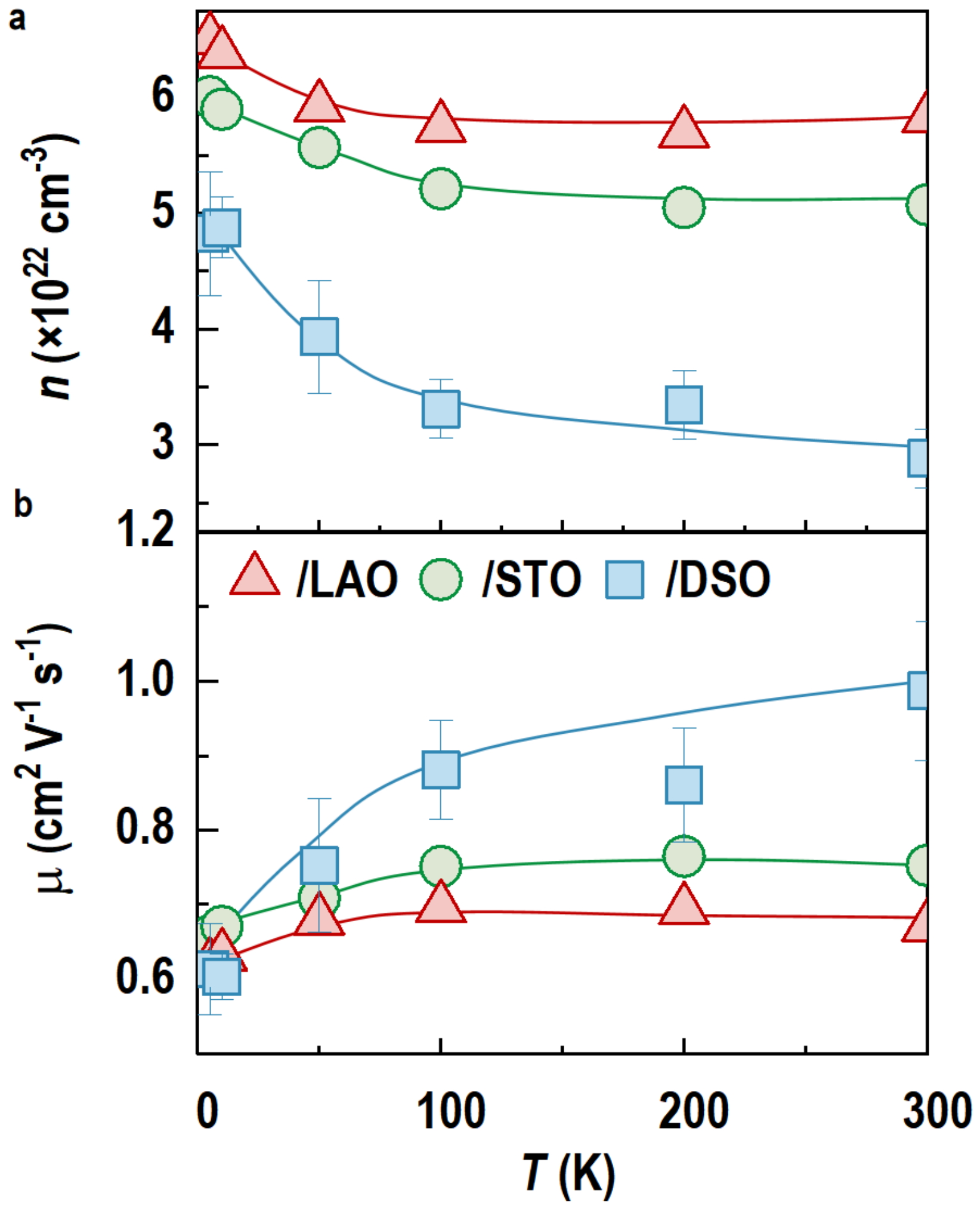

**fig. S14. Carrier densities (*n*) and Hall mobilities (*μ*) of $Ni_3InN$** thin films grown on $LaAlO_3$ (red), $SrTiO_3$ (green) and $DyScO_3$ (blue) substrates as functions of temperature.